\documentclass[11pt]{article}
\usepackage[numbers,sort&compress]{natbib}

\usepackage[a4paper,margin=1in]{geometry}
\usepackage{amsmath,amssymb,amsfonts}
\usepackage{booktabs}
\usepackage{tabularx}
\usepackage{graphicx}
\usepackage{hyperref}
\usepackage{xcolor}
\usepackage{natbib}
\usepackage{enumitem}
\usepackage{caption}
\usepackage{array}
\usepackage{placeins}
\usepackage{float}

\usepackage{makecell}

\hypersetup{
  colorlinks=true,
  linkcolor=blue,
  citecolor=blue,
  urlcolor=blue
}

\title{Transfer Learning Across Policy Regimes in Adaptive Multi-Agent Systems}

\author{Roberto Garrone\\
Open University of Cyprus\\
\texttt{roberto.garrone@st.ouc.ac.cy}}

\date{}

\begin{document}

\maketitle

\begin{abstract}
	Policy models often assume that the relationship between a policy instrument and its outcome remains stable across institutional conditions. In adaptive socio-technical systems this assumption may fail: regulatory change can alter incentives, agents can respond strategically, and the mapping from policy variables to aggregate outcomes can change. This paper studies such regime change as a transfer-learning problem in adaptive multi-agent systems. A policy regime is represented as a learning problem induced by an observable input distribution and a target function mapping policy variables to outcomes. We compare a blank-slate learner that searches a flexible hypothesis class in the new regime with a transfer learner whose effective hypothesis class is restricted by structural knowledge from the previous regime. Transfer is beneficial when this restriction preserves the new target function while reducing effective complexity; it is harmful when the restriction excludes the new target and creates misspecification. A stylized emissions-regulation experimental environment and a dynamic ABM robustness experiment support the claim. When the target regime preserves an affine monotone tax--emissions relation, transfer improves empirical small-sample performance. When the target regime introduces a threshold break, the same transferred structure produces negative transfer: held-out error remains high, online prediction generates more mistakes, and repeated online streams show larger cumulative and final-window error under misspecification. The contribution is methodological: previous regulatory experience should be reused when it captures stable structural invariants, but treated cautiously when policy change alters the policy--outcome relationship.
	
\end{abstract}

\section{Introduction}

Learning in real-world policy environments often occurs under changing institutional, technological, and regulatory conditions. Regulators and analysts do not observe passive systems. They intervene in environments populated by heterogeneous agents who interpret rules, respond to incentives, adapt their behavior, and interact with one another. These responses can feed back into the environment and change the relationship between policy instruments and aggregate outcomes. Environmental regulation, energy policy, urban planning, epidemiological intervention, and financial supervision all exhibit this general structure: a policy is introduced, agents respond, the system changes, and subsequent policy decisions are made using data generated under previous interventions.

Such environments are naturally described as complex adaptive systems. Aggregate patterns emerge from heterogeneous interacting components rather than from a single centralized mechanism. Agent-based models (ABMs) are important in this context because they represent heterogeneous agents, bounded rationality, local interaction, and endogenous feedback explicitly \citep{epstein1999,gilbert2008,tesfatsionjudd2006,conte2014}. They are therefore well suited for studying policy interventions in adaptive socio-technical systems.

Classical learning theory begins from different assumptions. In the Probably Approximately Correct (PAC) framework, examples are usually assumed to be drawn from a stable distribution and the target concept is fixed \citep{valiant1984,kearns1994}. These assumptions make it possible to derive sample-complexity guarantees. One can ask how many observations are required for a learner to identify a hypothesis with error at most $\epsilon$ and confidence at least $1-\delta$. However, these assumptions become problematic when the environment changes in response to policy intervention.

Policy learning often proceeds as if the policy--outcome relationship is stable. A regulator may learn that a higher carbon tax tends to reduce emissions, that a congestion charge reduces traffic, or that a subsidy increases technology adoption. This prior knowledge can be valuable when a new regime preserves the same structure. It can also be misleading when the new regime changes how agents respond. Firms may adopt new technology after a threshold, delay compliance, relocate production, or respond strategically to expected future rules. In these cases, the relationship learned under one regime may no longer describe the next.

This paper studies transfer learning across policy regimes in adaptive multi-agent systems. A policy regime is treated as a configuration of rules, incentives, constraints, and institutional conditions that shapes the mapping from policy variables to system outcomes. When the regime changes, the learning problem may change as well. The central question is: \emph{Under what structural conditions does prior knowledge learned under policy regime A help or harm learning after transition to policy regime B?}

The paper compares two learners. A blank-slate learner enters the new regime and searches over a flexible hypothesis class. A transfer learner reuses structural knowledge from the previous regime by restricting its effective hypothesis class. This is not treated as a warm start or initialization trick. Transfer is modeled as structural inductive bias. It helps when the target function induced by the new regime remains inside the restricted class. It harms when the new regime moves the target outside that class.

The contribution is threefold. First, the paper formulates policy regimes as learning problems of the form $(D_R,f_R)$, where $D_R$ is the distribution of observable policy inputs and $f_R$ is the target policy--outcome mapping induced by regime $R$. Second, it identifies structural-property similarity as the key condition for useful transfer: prior knowledge should restrict the learner only when the new target preserves the relevant structural invariant. Third, it reports a controlled emissions-regulation experiment and a dynamic ABM robustness experiment showing both positive and negative transfer. In a structurally similar target regime, transfer reduces small-sample error. In a structural-break target regime, transfer creates persistent misspecification and much higher online mistake counts. The paper clarifies when previous regulatory knowledge should be reused and when it should be treated as a potentially harmful inductive bias. It does not solve optimal policy control, nor does it claim to model a real emissions sector. 

\paragraph{Relation to companion work.}
This paper is distinct from related work on adaptive policy-regime comparison in ABMs. That work asks whether static and adaptive controller assumptions generate structurally distinguishable trajectories around a regulatory constraint. The present paper instead studies transfer across policy-regime transitions as a learning problem: previous regulatory knowledge restricts the effective hypothesis class of a learner in the target regime. The emissions setting is used as a transparent benchmark, but the learners, evaluation metrics, and central claim concern positive and negative transfer under structural similarity or structural break.

\section{Related Work}

\subsection{Computational learning theory}

Computational learning theory provides the formal language for analyzing how target functions can be learned from examples. In the PAC framework introduced by \citet{valiant1984} and developed by \citet{kearns1994}, a learner receives examples and attempts to identify a hypothesis that approximates the target with bounded error and high confidence. Sample complexity depends on the complexity of the hypothesis class. Learning over a smaller or more structured class generally requires fewer observations than learning over a larger class, provided that the true target remains inside the smaller class.

This framework is directly relevant because transfer learning can be interpreted as a restriction of the effective hypothesis space. If knowledge from regime A restricts the learner from a large class $H$ to a smaller class $H_S$, then learning under regime B may require fewer samples if $f_B \in H_S$. If $f_B \notin H_S$, the learner becomes misspecified.

The standard PAC framework is less directly equipped to handle policy-induced non-stationarity. In adaptive policy systems, interventions may alter the mechanisms that generate outcomes. This paper therefore uses PAC-style sample-complexity reasoning as a formal lens, while using held-out test error, online prediction loss, and mistake counts as empirical adaptation measures.

\subsection{Transfer learning and negative transfer}

Transfer learning studies how knowledge acquired in one task affects performance in another related task \citep{pan2010,weiss2016}. In reinforcement learning, transfer may involve reusing policies, value functions, options, models, representations, or exploration strategies \citep{taylor2009}. Recent work also studies transfer under shifted dynamics and cross-domain policy transfer \citep{chen2024,zhao2024,nakamoto2024}.

A central problem in transfer learning is negative transfer. Prior knowledge is useful only when source and target tasks share relevant structure. If the prior induces an invalid bias, it can slow adaptation or increase error. This paper applies that insight to regulatory regimes. The source and target are not arbitrary task domains. They are policy regimes induced by institutional and incentive conditions. A regime transition is therefore not merely a data shift; it can be a policy-induced change in the target function.

\subsection{Concept drift and policy-induced non-stationarity}

Concept drift research studies situations in which the relationship between inputs and outputs changes over time \citep{gama2014}. Methods include drift detection, sliding windows, model updating, ensembles, and adaptive retraining. The central concern is that a model trained on past data may become inaccurate when the data-generating process changes.

The present paper shares this concern but differs in its interpretation of drift. In many concept-drift settings, drift is treated as exogenous. In policy-oriented multi-agent systems, the change may be endogenous to regulatory intervention. The regulator changes incentives or constraints, agents respond, and the policy--outcome relation changes. This is better described as policy-induced target shift.

\subsection{Agent-based modeling and policy regimes}

ABMs are widely used to study policy interventions in complex systems because they allow aggregate patterns to emerge from heterogeneous agent behavior and interaction \citep{epstein1999,gilbert2008,tesfatsionjudd2006}. Socio-technical transitions research similarly emphasizes that technologies, institutions, markets, and user practices co-evolve \citep{geels2002,geels2016}. Policy-transfer research studies how governments learn from previous policies or other jurisdictions \citep{dolowitz2000}. These literatures support the substantive intuition that regimes matter.

However, policy-oriented ABMs are not usually framed as environments for analyzing the sample complexity or mistake behavior of learners transferring across regulatory regimes. The gap addressed here is the absence of a learning-theoretic account of how prior regulatory knowledge affects adaptation when policy interventions change the target function faced by the learner.

\section{Problem Formulation}

Consider a multi-agent system evolving in discrete time. Let $s_t$ denote the system state, $X_t$ the vector of agent actions, $P_t$ a policy parameter or vector of policy parameters controlled by the regulator, and $\zeta_t$ exogenous shocks. The system evolves according to
\begin{equation}
  s_{t+1}=F(s_t,X_t,P_t,\zeta_t).
\end{equation}
The full dynamics $F$ may include agent behavior, interaction effects, environmental constraints, and stochastic shocks. The learner does not directly observe or know $F$. Instead, it observes a policy-relevant outcome
\begin{equation}
  y_t=\Phi(s_t),
\end{equation}
where $\Phi(\cdot)$ maps states to measurable indicators such as aggregate emissions, compliance, welfare, or stability.

The learner is a policy-learning component. It may represent a regulator, analyst, or decision-support system attempting to infer the relationship between policy parameters and observable outcomes. The learner does not optimize policy in the present formulation. It learns a predictive mapping from policy parameters to outcomes. This restriction keeps the paper focused on transfer across regimes rather than optimal control.

For a given policy parameter $P$, the reduced-form expected outcome under a regime can be written as
\begin{equation}
  f(P)=\mathbb{E}[\Phi(s_T)\mid P].
\end{equation}
This expression summarizes the aggregate effect of policy parameter $P$ through the underlying multi-agent dynamics.

\subsection{Online prediction setting}

At each time step $t$, the learner observes a policy input $P_t$ and predicts the resulting outcome using a hypothesis $h_t$:
\begin{equation}
  \hat{y}_t=h_t(P_t), \qquad h_t\in H.
\end{equation}
After observing the realized outcome $y_t$, the learner updates according to
\begin{equation}
  h_{t+1}=U(h_t,P_t,y_t).
\end{equation}
The policy input $P_t$, hypothesis $h_t$, and realized outcome $y_t$ have distinct roles. $P_t$ is the regulatory setting whose consequences are to be predicted. $h_t$ is the learner's current predictive model. $y_t$ is the outcome generated by the multi-agent system. Performance is evaluated by cumulative prediction loss
\begin{equation}
  L_T=\sum_{t=1}^{T}\ell(h_t(P_t),y_t),
\end{equation}
and by mistake count
\begin{equation}
  M_T=\sum_{t=1}^{T}\mathbb{I}(|h_t(P_t)-y_t|>\rho),
\end{equation}
where $\rho$ is a prediction-error threshold.

\subsection{Policy regimes as learning problems}

A policy regime $R$ is represented as a learning problem
\begin{equation}
  R=(D_R,f_R),
\end{equation}
where $D_R$ is the distribution of observable policy inputs and $f_R$ is the target mapping from policy inputs to outcomes under that regime. The regime includes institutional rules, incentives, technological context, enforcement constraints, and other conditions that shape agent behavior. A regime-specific reduced-form mapping is therefore
\begin{equation}
  f_R(P)=\mathbb{E}[\Phi(s_T)\mid P,\pi_R],
\end{equation}
where $\pi_R$ denotes the regime configuration.

A transition from regime A to regime B can then be written as
\begin{equation}
  (D_A,f_A)\rightarrow(D_B,f_B).
\end{equation}
This separates pure target-function shift, pure distribution shift, and the more general adaptive-policy case:
\begin{equation}
  D_A\neq D_B, \qquad f_A\neq f_B.
\end{equation}
The central transfer problem is that a learner may possess useful knowledge about $f_A$, but after the regime transition it must learn $f_B$.

\section{Structural Similarity and Transfer}

A regime transition can alter the target function, the input distribution, or both. But numerical difference alone does not determine whether transfer should help. A higher carbon tax may change the level of emissions while preserving a monotone decreasing tax--emissions relation. Conversely, a regime may appear distributionally similar while changing the functional structure in a policy-relevant way. The relevant question is whether the structural constraints learned under the source regime remain valid under the target regime.

Let $H$ denote a flexible hypothesis class and let $S$ denote structural constraints learned or assumed from regime A. These constraints define a restricted subclass
\begin{equation}
  H_S=\{h\in H: h \text{ satisfies } S\}.
\end{equation}
Regimes A and B are structurally similar with respect to $S$ when
\begin{equation}
  f_A,f_B\in H_S\subseteq H.
\end{equation}
In this case, transferred knowledge restricts the learner to a smaller class that still contains the new target. If the restricted class has lower effective complexity, transfer can reduce the number of examples required for learning.

By contrast, transfer becomes harmful when
\begin{equation}
  f_B\notin H_S, \qquad f_B\in H.
\end{equation}
Then the transfer learner is misspecified. Even with abundant target-regime data, it cannot recover $f_B$ while constrained to $H_S$. The blank-slate learner may require more data initially, but it can represent the target if it searches over $H$.

\paragraph{Proposition 1: transfer under structural similarity.}
Let $H_S\subseteq H$ and suppose $f_B\in H_S$. If the effective complexity of $H_S$ is lower than that of $H$, then learning over $H_S$ requires no more samples than learning over $H$, up to the usual complexity-dependent terms in PAC-style bounds.

\paragraph{Proposition 2: negative transfer under structural dissimilarity.}
Let $H_S\subseteq H$ and suppose $f_B\notin H_S$ while $f_B\in H$. Then a transfer learner restricted to $H_S$ is misspecified in regime B, while a blank-slate learner searching over $H$ may still learn $f_B$.

These propositions state the central logic rather than a new learning-theoretic bound. In the finite-class case, the standard PAC bound scales as
\begin{equation}
  m_H(\epsilon,\delta)=O\left(\frac{1}{\epsilon}\left(\log |H|+\log\frac{1}{\delta}\right)\right),
\end{equation}
while a learner restricted to $H_S$ satisfies
\begin{equation}
  m_{H_S}(\epsilon,\delta)=O\left(\frac{1}{\epsilon}\left(\log |H_S|+\log\frac{1}{\delta}\right)\right).
\end{equation}
Thus, when $|H_S|<|H|$ and $f_B\in H_S$, transfer can reduce the empirical number of target-regime observations required to reach the chosen held-out error criterion. When $f_B\notin H_S$, the bound is irrelevant because the learner is biased toward the wrong class.

The PAC expressions above are used only as a formal intuition for the effect of restricting a hypothesis class. They are not claimed as exact sample-complexity bounds for the implemented estimators. The empirical learners used below are continuous least-squares regressors, and the blank-slate learner searches a finite grid of threshold locations. Consequently, the experiments report empirical sample requirements: the smallest tested training size for which the fitted learner reaches a predefined held-out error criterion with the required empirical frequency across repetitions. The term sample complexity is therefore reserved for the PAC-style motivation, while empirical results are interpreted as finite-experiment evidence about sample efficiency under the specified learners, grids, noise levels, and target regimes.

\section{Emissions-Regulation Experimental Environment}

The stylized emissions-regulation ABM is not intended as a realistic environmental-policy model. Its purpose is to provide a controlled setting in which positive and negative transfer can be compared.

There are $N=100$ firms. Firm $i$ has baseline emissions $b_i$, tax responsiveness $\eta_i$, and optional threshold response $\theta_i$. For a carbon tax $\lambda$, emissions are generated by
\begin{equation}
  e_i(\lambda)=\max\{0,b_i-\eta_i\lambda+\theta_i(\lambda-\tau)_+ + \varepsilon_i\},
\end{equation}
where $(x)_+=\max\{x,0\}$ and $\tau$ is the threshold. Aggregate emissions are
\begin{equation}
  E(\lambda)=\sum_{i=1}^{N}e_i(\lambda).
\end{equation}
The reduced-form target is approximately
\begin{equation}
  f_R(\lambda)=\alpha_R-\beta_R\lambda+\gamma_R(\lambda-\tau_R)_+.
\end{equation}
The parameters $(\alpha_R,\beta_R,\gamma_R,\tau_R)$ define the regime.

The source regime A is affine monotone:
\begin{equation}
  \alpha_A=100,\quad \beta_A=2.0,\quad \gamma_A=0,\quad \tau_A=10.
\end{equation}
The first target regime B1 is structurally similar:
\begin{equation}
  \alpha_{B1}=105,\quad \beta_{B1}=2.8,\quad \gamma_{B1}=0,\quad \tau_{B1}=10.
\end{equation}
It changes the intercept and slope but preserves the affine monotone structure.

The second target regime B2 introduces a threshold break:
\begin{equation}
  \alpha_{B2}=105,\quad \beta_{B2}=2.8,\quad \gamma_{B2}=4.0,\quad \tau_{B2}=10.
\end{equation}
For $\lambda>10$, the positive threshold term changes the slope of the tax--emissions relation. This represents a structural response such as rebound behavior, relocation, delayed compliance, or strategic adjustment that weakens the emissions-reduction effect after a policy threshold.

\subsection{Hypothesis classes}

The blank-slate learner searches over a flexible piecewise-linear threshold class:
\begin{equation}
  H=\{h(\lambda)=a-b\lambda+c(\lambda-\tau)_+ : b\geq 0,\ \tau\in\mathcal{T}\}.
\end{equation}
In the experiment, $\mathcal{T}$ is a finite grid of 13 threshold values between 4 and 16.

The transfer learner searches over the restricted affine monotone class:
\begin{equation}
  H_S=\{h(\lambda)=a-b\lambda : b\geq 0\}.
\end{equation}
This class captures the structural knowledge supported by regime A: emissions decrease approximately linearly with the tax. It is appropriate for B1 but misspecified for B2.

Both learners are fitted using empirical risk minimization with squared loss. The transfer learner is not given target-regime outcomes in advance. It simply uses a restricted hypothesis class when fitting target-regime samples. This isolates the effect of structural transfer from warm-start effects.

\subsection{Dynamic ABM robustness experiment}

The reduced-form experiment isolates the learning-theoretic mechanism by specifying the target functions directly. This makes the source of positive and negative transfer transparent, but it leaves open whether the same pattern appears when the policy--outcome mapping is generated by adaptive agents and policy feedback. To address this, we add a dynamic ABM robustness experiment.

In this experiment, outcomes are generated by \(N\) heterogeneous firms. Firm \(i\)'s emissions at time \(t\) are

\[
e_{i,t}
=
\max\{0,\,
b_i
-
\eta_{i,t}\lambda_t
+
\theta_i(\lambda_t-\tau)_+
+
\varepsilon_{i,t}
\}.
\]

Here, \(b_i\) is baseline emissions, \(\eta_{i,t}\) is the firm's time-varying responsiveness to the carbon tax, \(\lambda_t\) is the realized tax at time \(t\), \(\theta_i\) controls the strength of a threshold response, \(\tau\) is the threshold, and \(\varepsilon_{i,t}\) is an idiosyncratic shock. Aggregate emissions are

\[
E_t=\sum_{i=1}^{N}e_{i,t}.
\]

Firms adapt their responsiveness over time. Adaptation depends on cap pressure and on local peer effects:

\[
\eta_{i,t+1}
=
\mathrm{clip}
\left(
\eta_{i,t}
+
\kappa_i\frac{(E_t-C)_+}{C}
+
\omega(\bar{\eta}_{\mathcal{N}(i),t}-\eta_{i,t}),
0,
\eta_{\max}
\right).
\]

The first adjustment term increases responsiveness when aggregate emissions exceed the cap \(C\). The second term represents peer influence: \(\bar{\eta}_{\mathcal{N}(i),t}\) is the average responsiveness of firm \(i\)'s local neighbors, and \(\omega\) controls the strength of imitation or diffusion. The clipping operator keeps responsiveness within the feasible interval \([0,\eta_{\max}]\).

The regulator also updates the realized tax endogenously around a base policy input \(\lambda^{base}\):

\[
\lambda_{t+1}
=
\mathrm{clip}
\left(
\lambda^{base}
+
\gamma_P(E_t-C),
\lambda_{\min},
\lambda_{\max}
\right).
\]

Thus, the learner observes the base policy input \(\lambda^{base}\), but the actual trajectory is generated by feedback between emissions, firm adaptation, and policy adjustment.

\begin{table}[H]
	\centering
	\caption{Dynamic ABM robustness experiment: main parameter settings.}
	\label{tab:dynamic-abm-settings}
	\begin{tabular}{ll}
		\toprule
		Parameter & Value \\
		\midrule
		Number of firms \(N\) & 100 \\
		Simulation horizon \(T\) & 80 \\
		Burn-in \(B\) & 20 \\
		Post-burn outcome & Mean aggregate emissions over \(t \geq B\) \\
		Emissions cap \(C\) & 75 \\
		Base-tax distribution & \(\lambda^{base} \sim \mathrm{Unif}(0,20)\) \\
		Realized-tax bounds & \([\lambda_{\min},\lambda_{\max}] = [0,25]\) \\
		Policy-feedback coefficient \(\gamma_P\) & 0.015 \\
		Baseline emissions \(b_i\) & \(b_i \sim \mathrm{Unif}(0.6,1.4)\) \\
		Initial responsiveness \(\eta_{i,0}\) & \(\eta_{i,0} \sim \mathrm{Unif}(0.010,0.030)\times \eta_{\mathrm{scale}}\) \\
		Responsiveness bound \(\eta_{\max}\) & 0.080 \\
		Adaptation gain \(\kappa_i\) & \(\kappa_i \sim \mathrm{Unif}(0.75,1.25)\times \kappa_R\) \\
		Peer-effect strength \(\omega\) & 0.015 \\
		Network structure & Ring network, two nearest neighbors \\
		Noise standard deviation & 0.03 \\
		Threshold \(\tau\) & 10 \\
		Threshold grid for blank learner & 13 values in \([4,16]\) \\
		Dynamic sample-size repetitions & 50 \\
		Dynamic sample sizes & \(n \in \{10,20,50,100,200\}\) \\
		Dynamic test-set size & 400 \\
		Sample-size experiment seed & 20260503 \\
		Online horizon & 100 \\
		Online base-tax distribution & \(\lambda^{base}_t \sim \mathrm{Unif}(0,20)\) \\
		Online mistake threshold \(\rho\) & 8 \\
		Online experiment seed & 303030 \\
		Online repeated streams & 50 \\
		Online base seed & 303030 \\
		Online seed stride & 7919 \\
		Post-warm-up window & t = 10,\ldots,99 \\
		Final-window MSE & t = 75,\ldots,99 \\
		\bottomrule
	\end{tabular}
\end{table}

The induced target function for regime \(R\) is defined as the expected aggregate emissions generated by the dynamic system (after the initial transient phase):

\[
f_R(\lambda^{base})
=
\mathbb{E}
\left[
\frac{1}{T-B}\sum_{t=B+1}^{T}E_t
\mid
\lambda^{base},\pi_R
\right],
\]

where \(T\) is the simulation horizon, \(B\) is the initial transient period, and \(\pi_R\) denotes the regime configuration. This definition keeps the learning problem aligned with the reduced-form experiment: the learner still estimates a mapping from a policy input to an aggregate outcome, but the mapping is now induced by dynamic heterogeneous agents and feedback rather than specified directly.

We define three regimes. The source regime \(A\) has no threshold response and a baseline level of adaptive responsiveness:

\[
A:\quad \theta_i=0,\quad \kappa_i=\kappa_A.
\]

The structurally similar target regime \(B_1\) preserves the same qualitative tax--emissions relation but changes the strength of adaptation:

\[
B_1:\quad \theta_i=0,\quad \kappa_i>\kappa_A.
\]

The structural-break target regime \(B_2\) introduces a threshold response for some firms once the realized tax exceeds \(\tau\):

\[
B_2:\quad \theta_i>0 \text{ for some firms when } \lambda_t>\tau.
\]

Regime \(B_1\) is designed to preserve the affine monotone structure learned from the source regime, so transfer to the restricted affine class \(H_S\) should remain useful. Regime \(B_2\) changes the policy--outcome structure by introducing a threshold effect, so the same transferred restriction can become misspecified. The robustness experiment therefore tests whether the positive-transfer and negative-transfer pattern observed in the reduced-form experiment also appears when the target function is generated by an adaptive multi-agent process.

\begin{table}[H]
	\centering
	\caption{Dynamic ABM robustness experiment: regime-specific parameters.}
	\label{tab:dynamic-abm-regimes}
	\begin{tabular}{lllll}
		\toprule
		Regime & Interpretation & \(\eta_{\mathrm{scale}}\) & \(\kappa_R\) & Threshold response \(\theta_i\) \\
		\midrule
		\(A\) & Source affine-like regime & 1.00 & 0.0015 & \(\theta_i=0\) \\
		\(B_1\) & Structurally similar dynamic target & 1.25 & 0.0025 & \(\theta_i=0\) \\
		\(B_2\) & Threshold-break dynamic target & 1.25 & 0.0025 & \(\theta_i\sim\mathrm{Unif}(0.025,0.055)\) \\
		\bottomrule
	\end{tabular}
\end{table}

\section{Experimental Design}

We conduct two experiment families. The first uses directly specified reduced-form target functions to isolate the learning-theoretic mechanism. The second uses a dynamic ABM-generated target function as a robustness check. In each family, we run a sample-size experiment and an online adaptation experiment.

\paragraph{Sample-size experiment.}
For each target regime and learner, training sets of size
\begin{equation}
  n\in\{10,20,50,100,200\}
\end{equation}
are generated from the target regime. For each condition, both the reduced-form sample-size experiment and the dynamic ABM robustness experiment use \(R=50\) repetitions. Each fitted model is evaluated on a held-out target-regime test set with 400 observations. The reported metric is mean squared error (MSE), averaged over repetitions.

An empirical sample requirement is also computed for threshold $\epsilon=10$ and confidence target $1-\delta=0.9$. The requirement is the smallest $n$ for which the test MSE is below $\epsilon$ in at least 90\% of repetitions.

\paragraph{Online adaptation experiment.}
In the online experiment, each learner faces a sequence of 100 target-regime policy inputs. At each step it predicts using the model fitted on previous observations, observes the outcome, records squared loss, and refits using all data seen so far. A mistake is counted when absolute error exceeds $\rho = 8$. For the reduced-form experiment, we report final cumulative loss and final cumulative mistakes. For the dynamic ABM robustness experiment, we additionally repeat the online procedure over 50 independent streams and report mean mistake counts, median cumulative loss, median post-warm-up loss, and final-window MSE.

\section{Results}
The results are reported in two stages. Sections 7.1 and 7.2 present the reduced-form experiment, which isolates the learning-theoretic mechanism. Section 7.3 presents the dynamic ABM robustness experiment, which tests whether the same pattern appears when the target mapping is induced by adaptive heterogeneous firms and policy feedback.

\subsection{Reduced-form sample-size experiment}

Table~\ref{tab:sample-size} reports held-out MSE by sample size over 50 repetitions. The results show the expected conditional pattern. In \(B_1\), where the affine structure is preserved, the transfer learner has lower error at small sample sizes: at \(n=10\), its MSE is \(7.01 \pm 1.04\), compared with \(13.85 \pm 33.44\) for the blank-slate piecewise learner. The learners converge at larger sample sizes.

In \(B_2\), where the target contains a threshold break, the transfer learner performs poorly even with 200 samples. At \(n=200\), its MSE remains \(39.45 \pm 2.44\), compared with \(6.30 \pm 0.45\) for the blank-slate learner. The issue is not lack of data but misspecification: \(H_S\) cannot represent the target relation.

The finite-experiment sample requirement reinforces this interpretation. In \(B_1\), the transfer learner reaches the criterion at \(n=10\), while the blank-slate learner reaches it only at \(n=50\).

\begin{table}[H]
	\centering
	\caption{Held-out MSE by target regime, learner, and training sample size. Values are mean \(\pm\) standard deviation over 50 repetitions.}
	\label{tab:sample-size}
	\resizebox{\textwidth}{!}{
		\begin{tabular}{llrrrrr}
			\toprule
			Regime & Learner & \(n=10\) & \(n=20\) & \(n=50\) & \(n=100\) & \(n=200\) \\
			\midrule
			\(B_1\) affine similar & Blank \(H\) piecewise & \(13.85 \pm 33.44\) & \(10.69 \pm 16.30\) & \(6.63 \pm 0.65\) & \(6.33 \pm 0.64\) & \(6.20 \pm 0.44\) \\
			\(B_1\) affine similar & Transfer \(H_S\) affine & \(7.01 \pm 1.04\) & \(6.69 \pm 0.82\) & \(6.30 \pm 0.55\) & \(6.22 \pm 0.48\) & \(6.17 \pm 0.43\) \\
			\(B_2\) threshold break & Blank \(H\) piecewise & \(23.05 \pm 72.52\) & \(7.95 \pm 1.59\) & \(6.79 \pm 0.79\) & \(6.41 \pm 0.53\) & \(6.30 \pm 0.45\) \\
			\(B_2\) threshold break & Transfer \(H_S\) affine & \(57.68 \pm 29.18\) & \(43.88 \pm 5.49\) & \(41.01 \pm 2.78\) & \(40.15 \pm 2.80\) & \(39.45 \pm 2.44\) \\
			\bottomrule
		\end{tabular}
	}
\end{table}

The empirical sample requirement reinforces this interpretation. In \(B_1\), the transfer learner reaches the criterion at \(n=10\), while the blank-slate learner reaches it only at \(n=50\). In \(B_2\), the blank-slate learner reaches the criterion at \(n=20\), while the transfer learner does not reach it for any tested sample size.

\begin{table}[H]
	\centering
	\caption{Empirical sample requirement for MSE threshold \(\epsilon=10\) and confidence target \(1-\delta=0.9\).}
	\label{tab:sample-requirement}
	\begin{tabular}{lll}
		\toprule
		Regime & Learner & Smallest \(n\) meeting criterion \\
		\midrule
		\(B_1\) affine similar & Blank \(H\) piecewise & 50 \\
		\(B_1\) affine similar & Transfer \(H_S\) affine & 10 \\
		\(B_2\) threshold break & Blank \(H\) piecewise & 20 \\
		\(B_2\) threshold break & Transfer \(H_S\) affine & not reached \\
		\bottomrule
	\end{tabular}
\end{table}

\subsection{Reduced-form online adaptation}

Table~\ref{tab:online-results} reports online adaptation results. In B1, the two learners behave similarly. The transfer learner has slightly lower cumulative loss and the same number of mistakes. This supports the claim that transfer is beneficial or at least harmless when the transferred structure remains valid.

In B2, the result changes sharply. The blank-slate learner makes 5 mistakes, while the transfer learner makes 30. The transfer learner also accumulates substantially higher loss. This is the negative-transfer case: the restricted affine class cannot represent the threshold break, so the learner is biased toward the wrong model class.

\begin{table}[H]
	\centering
	\caption{Online adaptation results over horizon \(T=100\) with mistake threshold \(\rho=8\).}
	\label{tab:online-results}
	\begin{tabular}{llrr}
		\toprule
		Regime & Learner & Final cumulative loss & Mistakes \\
		\midrule
		\(B_1\) affine similar & Blank \(H\) piecewise & 9210.92 & 2 \\
		\(B_1\) affine similar & Transfer \(H_S\) affine & 9142.76 & 2 \\
		\(B_2\) threshold break & Blank \(H\) piecewise & 8889.93 & 5 \\
		\(B_2\) threshold break & Transfer \(H_S\) affine & 12441.90 & 30 \\
		\bottomrule
	\end{tabular}
\end{table}

\subsection{Dynamic ABM robustness}

The dynamic ABM robustness experiment reproduces the main sample-size pattern from the reduced-form experiment. Table~\ref{tab:dynamic-sample-size} reports held-out MSE for the dynamic ABM-generated target functions. In the structurally similar regime \(B_1\), the affine transfer learner has lower empirical small-sample error than the blank-slate piecewise learner: at \(n=10\), its MSE is \(8.34 \pm 2.46\), compared with \(14.29 \pm 24.47\) for the blank-slate learner. The flexible learner becomes slightly better at larger sample sizes, but the transfer learner provides the expected small-sample advantage.

In the threshold-break regime \(B_2\), the pattern reverses. The blank-slate learner over the piecewise class learns the threshold response, while the affine transfer learner remains misspecified. Even at \(n=200\), the transfer learner has MSE \(14.35 \pm 0.25\), compared with \(1.75 \pm 0.07\) for the blank-slate learner. The empirical sample-requirement table reinforces this result: under the threshold \( \epsilon = 10 \) and confidence target \(1-\delta=0.9\), transfer reaches the criterion at \(n=10\) in \(B_1\), but does not reach it in \(B_2\).

\begin{table}[H]
	\centering
	\caption{Dynamic ABM robustness experiment: held-out MSE by target regime, learner, and training sample size. Values are mean $\pm$ standard deviation over repetitions.}
	\label{tab:dynamic-sample-size}
	\resizebox{\textwidth}{!}{
		\begin{tabular}{llrrrrr}
			\toprule
			Regime & Learner & \(n=10\) & \(n=20\) & \(n=50\) & \(n=100\) & \(n=200\) \\
			\midrule
			\(B_1\) affine-similar dynamic & Blank \(H\) piecewise & \(14.29 \pm 24.47\) & \(7.26 \pm 1.64\) & \(6.31 \pm 0.50\) & \(6.16 \pm 0.18\) & \(6.05 \pm 0.11\) \\
			\(B_1\) affine-similar dynamic & Transfer \(H_S\) affine & \(8.34 \pm 2.46\) & \(7.28 \pm 0.54\) & \(6.79 \pm 0.25\) & \(6.71 \pm 0.19\) & \(6.61 \pm 0.09\) \\
			\(B_2\) threshold-break dynamic & Blank \(H\) piecewise & \(4.16 \pm 6.08\) & \(2.61 \pm 1.29\) & \(1.96 \pm 0.27\) & \(1.85 \pm 0.18\) & \(1.75 \pm 0.07\) \\
			\(B_2\) threshold-break dynamic & Transfer \(H_S\) affine & \(19.27 \pm 7.24\) & \(16.67 \pm 3.05\) & \(14.86 \pm 0.73\) & \(14.47 \pm 0.43\) & \(14.35 \pm 0.25\) \\
			\bottomrule
		\end{tabular}
	}
\end{table}

\begin{table}[H]
	\centering
	\caption{Dynamic ABM robustness experiment: empirical sample requirement for held-out MSE threshold $\epsilon = 10$ and empirical frequency target $1-\delta = 0.9$. The reported value is the smallest tested training size $n$ for which the criterion is met in at least 90\% of repetitions; it is not a theoretical sample-complexity bound.}
	\label{tab:dynamic-sample-requirements}
	\begin{tabular}{lll}
		\toprule
		Regime & Learner & Smallest \(n\) meeting criterion \\
		\midrule
		\(B_1\) affine-similar dynamic & Blank \(H\) piecewise & 20 \\
		\(B_1\) affine-similar dynamic & Transfer \(H_S\) affine & 10 \\
		\(B_2\) threshold-break dynamic & Blank \(H\) piecewise & 10 \\
		\(B_2\) threshold-break dynamic & Transfer \(H_S\) affine & not reached \\
		\bottomrule
	\end{tabular}
\end{table}

The online dynamic experiment is evaluated over repeated streams to avoid drawing conclusions from a single trajectory. Each stream uses matched policy-input and outcome sequences for both learners within the same regime, making the comparison paired within stream. We report mistake counts, median cumulative loss, post-warm-up loss after the first ten observations, and final-window MSE over the last 25 observations.

\begin{table}[H]
	\centering
	\scriptsize
	\setlength{\tabcolsep}{3pt}
	\renewcommand{\arraystretch}{1.18}
	\caption{Repeated dynamic ABM online adaptation diagnostics over 50 independent online streams. 
		Values report mean $\pm$ standard deviation for mistake counts and final-window MSE, and medians for cumulative squared loss measures. 
		Post-warm-up loss is computed after the first ten online observations; final-window MSE is computed over the last 25 observations.}
	\label{tab:dynamic-online-multistream}
	\begin{tabularx}{\textwidth}{@{}
			p{4cm}
			p{2.5cm}
			>{\centering\arraybackslash}p{1.75cm}
			>{\centering\arraybackslash}p{1.85cm}
			>{\centering\arraybackslash}p{1.95cm}
			>{\centering\arraybackslash}p{1.85cm}
			@{}}
		\toprule
		Regime 
		& Learner 
		& \makecell{Mistakes\\mean $\pm$ sd} 
		& \makecell{Median\\cumulative\\loss} 
		& \makecell{Median\\post-warm-up\\loss} 
		& \makecell{Final-window\\MSE\\mean $\pm$ sd} \\
		\midrule
		
		\(B_1\) affine-similar dynamic 
		& Blank $H$ piecewise 
		& $2.34 \pm 1.04$ 
		& $933.44$ 
		& $554.75$ 
		& $5.75 \pm 1.97$ \\
		
		\(B_1\) affine-similar dynamic 
		& Transfer $H_S$ affine 
		& $1.58 \pm 0.84$ 
		& $825.25$ 
		& $571.08$ 
		& $6.13 \pm 1.79$ \\
		
		\(B_2\) threshold-break dynamic 
		& Blank $H$ piecewise 
		& $1.50 \pm 0.91$ 
		& $461.31$ 
		& $177.47$ 
		& $1.55 \pm 0.66$ \\
		
		\(B_2\) threshold-break dynamic 
		& Transfer $H_S$ affine 
		& $5.44 \pm 1.76$ 
		& $1723.76$ 
		& $1321.61$ 
		& $13.39 \pm 4.83$ \\
		\bottomrule
	\end{tabularx}
\end{table}

\begin{figure}[H]
	\centering
	\includegraphics[width=.8\textwidth]{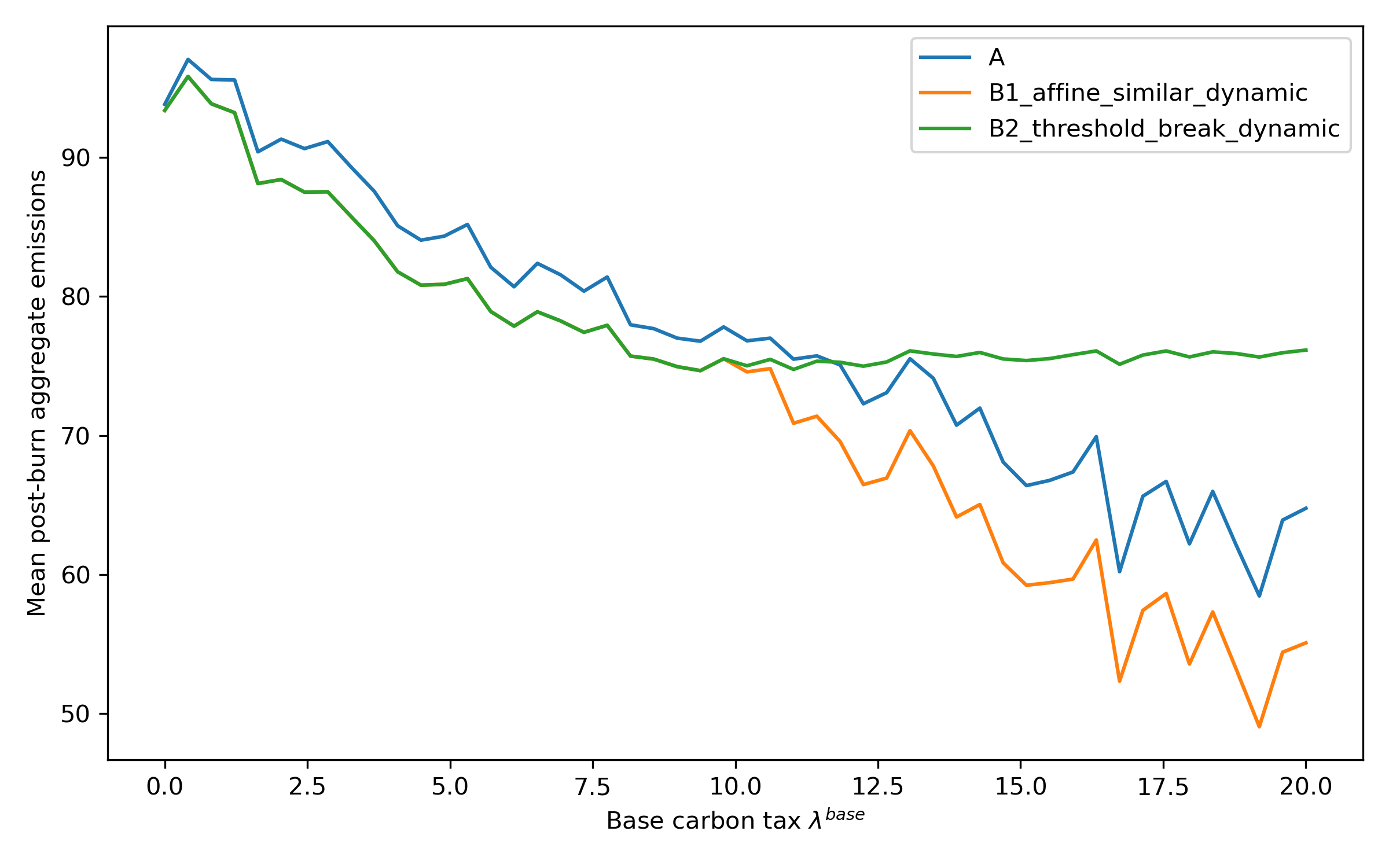}
	\caption{Dynamic ABM-induced policy--outcome mappings. The learner observes the base carbon tax \(\lambda^{base}\), while the realized trajectory is generated by heterogeneous firms, adaptive responsiveness, peer effects, and policy feedback. Regime \(B_1\) preserves a broadly monotone relation, whereas \(B_2\) introduces a threshold-induced flattening after high tax values.}
	\label{fig:dynamic-induced-functions}
\end{figure}

\begin{figure}[H]
	\centering
	\includegraphics[width=.8\textwidth]{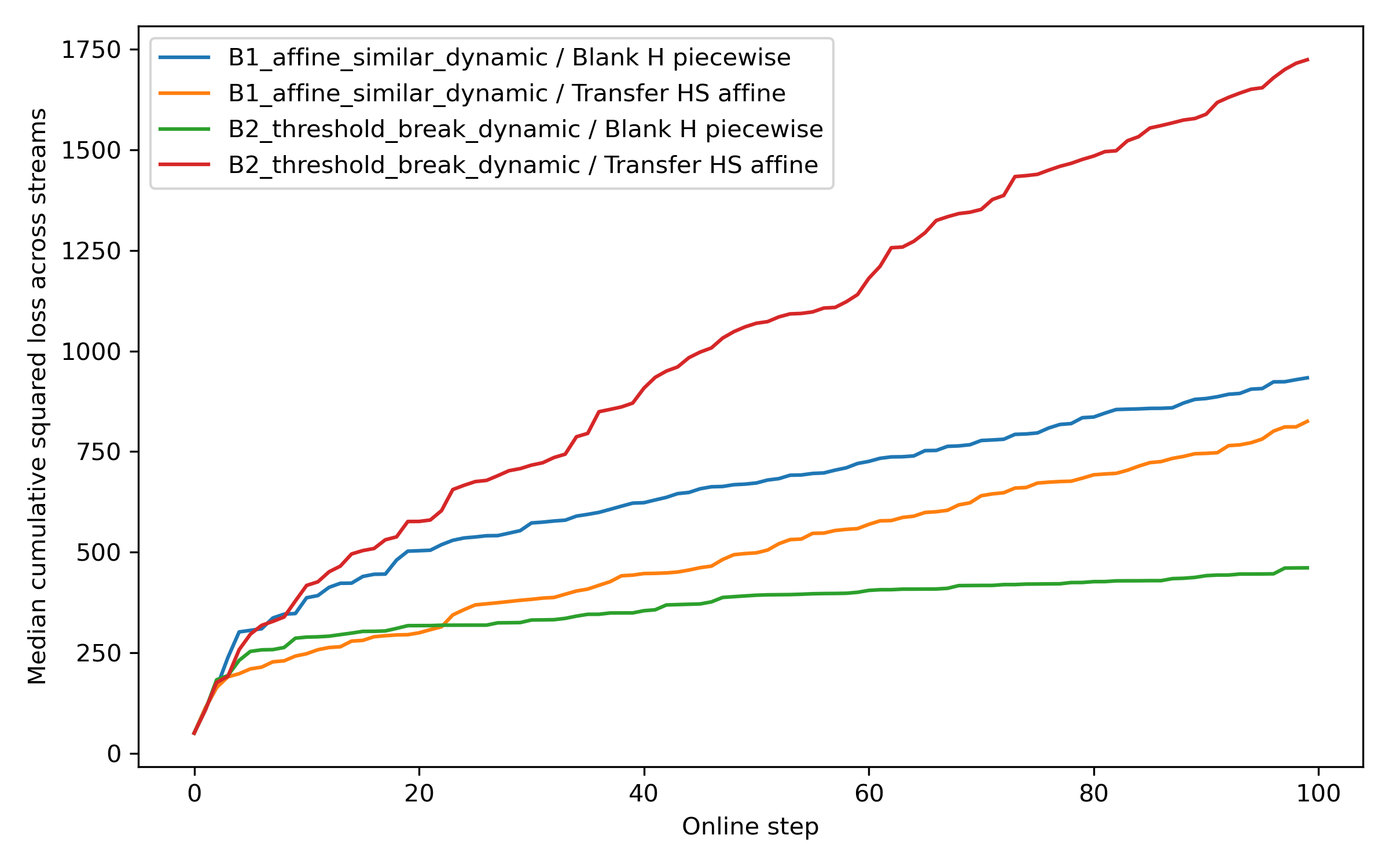}
	\caption{Median cumulative squared loss across 50 independent dynamic ABM online streams. In the structurally similar dynamic regime \(B_1\), the transfer learner remains broadly competitive with the blank-slate learner. In the threshold-break dynamic regime \(B_2\), the transfer learner accumulates substantially higher median loss, while the blank-slate learner adapts to the threshold structure and stabilizes at a much lower loss. The result supports the negative-transfer interpretation: when the target regime violates the transferred affine structure, the restricted learner remains persistently misspecified.}
	\label{fig:dynamic-online-median-loss}
\end{figure}

The repeated online experiment resolves the ambiguity visible in the single-stream diagnostic. In the structurally similar dynamic regime \(B_1\), the transfer learner remains broadly competitive with the blank-slate learner: it makes fewer mistakes on average and has lower median cumulative loss, although its final-window MSE is slightly higher. This is consistent with the interpretation that the affine restriction remains approximately valid when the target regime preserves the source-regime structure. In the threshold-break dynamic regime \(B_2\), the pattern is unambiguous. The affine transfer learner makes substantially more mistakes, accumulates much higher median cumulative and post-warm-up loss, and retains a final-window MSE almost one order of magnitude larger than that of the blank-slate learner. Figure~\ref{fig:dynamic-online-median-loss} confirms the same pattern visually: in \(B_2\), the transfer learner accumulates much higher median squared loss throughout the online horizon, whereas the blank-slate learner benefits from its more flexible threshold class. Thus, the negative-transfer result is not driven by a transient anomaly in one online trajectory, but reflects persistent misspecification of the transferred affine structure when the target regime introduces a threshold response.

\begin{figure}[H]
	\centering
	\includegraphics[width=.8\textwidth]{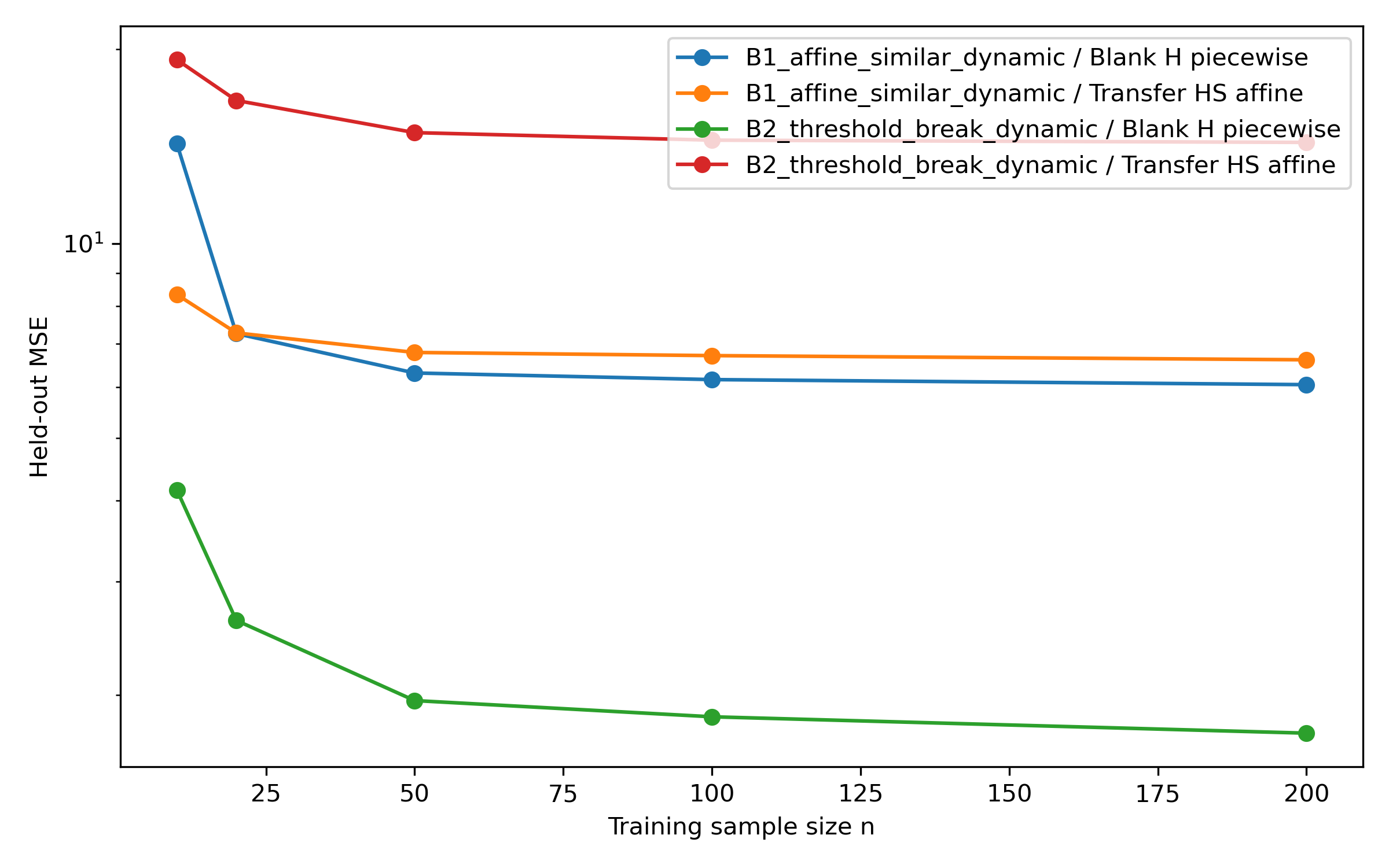}
	\caption{Held-out MSE in the dynamic ABM robustness experiment. Transfer improves small-sample performance in the structurally similar regime \(B_1\), but remains misspecified in the threshold-break regime \(B_2\). The y-axis is logarithmic.}
	\label{fig:dynamic-mse}
\end{figure}

Thus, the dynamic ABM robustness experiment confirms the sample-size and misspecification mechanism observed in the reduced-form experiment. Transfer remains useful or harmless when the structural invariant is preserved, but becomes harmful when the target regime moves outside the restricted hypothesis class.

\subsection{Interpretation}

The empirical evidence supports the theoretical logic. Transfer helps in the structurally similar regime because the restricted class contains the new target. The transfer learner avoids the high small-sample variance of the flexible piecewise class. In the structural-break regime, the same restriction becomes harmful. The affine transfer learner cannot represent the threshold response, so additional samples do not eliminate its error.

The 50-repetition reduced-form experiment reinforces the sample-efficiency interpretation. In the structurally similar regime \(B_1\), the transfer learner reaches the empirical MSE criterion with \(n=10\), while the blank-slate learner requires \(n=50\). In the threshold-break regime \(B_2\), the blank-slate learner reaches the criterion at \(n=20\), while the transfer learner never reaches it over the tested sample sizes.

The contrast is useful for policy learning. Previous regulatory experience should not be reused merely because it is available. It should be reused when it captures a stable structural property of the policy--outcome relationship. When policy change introduces thresholds, feedback loops, or strategic responses, transferred knowledge can become a source of inertia rather than an advantage.

\section{Discussion}

The results support three claims.

First, transfer across policy regimes is best understood as structural inductive bias. The prior regime does not simply provide data. It suggests restrictions on which policy--outcome mappings are plausible. These restrictions improve empirical sample efficiency only when they remain valid.

Second, structural similarity is more important than numerical similarity alone. A new regime may change intercepts and slopes while preserving the same affine monotone relation. In that case transfer is useful. Conversely, a threshold break can make a superficially familiar setting structurally different.

Third, negative transfer is a misspecification problem. In B2, the transfer learner receives more data but remains biased because its hypothesis class excludes the target. The blank-slate learner is less efficient initially but more robust to structural change because it searches a richer class.

The paper also clarifies the relation between transfer learning and concept drift. The issue is not merely that the data stream changes over time. The change is induced by a policy regime transition that modifies incentives and agent responses. This makes regime change a mechanism-level shift, not only a statistical perturbation.

\section{Limitations}

Both empirical settings are stylized. The reduced-form experiment isolates the learning-theoretic mechanism by specifying target functions directly. The dynamic ABM robustness experiment adds heterogeneous firms, adaptive responsiveness, peer effects, and endogenous policy feedback, but it remains a controlled simulation rather than a calibrated emissions-sector model. The empirical claim is therefore methodological rather than predictive.

The source-regime structure is imposed rather than discovered by a separate structure-learning algorithm. This is intentional: the experiments test whether a theoretically justified structural restriction helps or harms after a regime transition. Future work should study how such restrictions can be inferred from source-regime data, how reliable such inferred structures are, and when they should be relaxed after evidence of target-regime mismatch.

The learners are deliberately simple empirical-risk minimizers. More sophisticated learners could use regularization, Bayesian priors, online drift detection, ensembles, model selection between transferred and flexible classes, or selective-transfer mechanisms. These extensions are important, but the current design keeps the comparison transparent: the difference between the learners is the structural restriction itself, not algorithmic complexity.

The PAC-style analysis is illustrative. The empirical learners use continuous least-squares fitting with a finite threshold grid, so the finite-class bound should be read as motivating the class-restriction logic rather than as an exact finite hypothesis count for the implementation. The paper does not introduce a new sample-complexity theorem; it uses standard class-restriction reasoning to clarify a policy-learning problem.

The online results also illustrate the importance of repeated-stream evaluation. A single online trajectory can be misleading because cumulative squared loss is sensitive to early instability. For this reason, the dynamic online experiment is repeated over 50 independent streams. Under this repeated evaluation, the threshold-break regime shows consistent negative transfer across mistake counts, median cumulative loss, post-warm-up loss, and final-window MSE.

Finally, the learner predicts outcomes rather than optimizing policy. This is a deliberate restriction. The paper studies whether prior knowledge helps the learner understand a new regime, not whether it can choose the best policy. Extending the framework from prediction to adaptive policy choice is a natural next step.

\section{Conclusion}

This paper studied transfer learning across policy regimes in adaptive multi-agent systems. It framed regimes as learning problems \((D_R,f_R)\) and analyzed transfer as a restriction from a flexible hypothesis class \(H\) to a structurally informed subclass \(H_S\). The central claim is conditional: transfer is beneficial when the new regime preserves the structural invariant encoded by \(H_S\), and harmful when the new regime moves the target outside that class.

The reduced-form emissions-regulation experiment supports this claim in a transparent setting. In a structurally similar affine regime, transfer improves empirical small-sample performance: in the reduced-form experiment it reaches the empirical MSE criterion with \(n=10\), while the blank-slate learner requires \(n=50\). In a threshold-break regime, the same affine restriction produces persistent misspecification: held-out error remains high even as the target-regime sample size increases. Online, it is harmless in the structurally similar regime and harmful in the threshold-break regime by mistake count and cumulative loss.

The dynamic ABM robustness experiment strengthens the interpretation. When the policy--outcome mapping is generated by heterogeneous firms, adaptive responsiveness, peer effects, and endogenous policy feedback, the same sample-size pattern appears: transfer is useful in the structurally similar dynamic regime and harmful in the threshold-break dynamic regime. Repeated online diagnostics further support this interpretation. In the threshold-break dynamic regime, the restricted affine learner makes more mistakes, accumulates higher median cumulative and post-warm-up loss, and retains substantially higher final-window MSE than the blank-slate learner. This indicates persistent misspecification of the transferred structure rather than a trajectory-specific artifact.

The broader lesson is that previous regulatory experience should not be reused merely because it is available. It is valuable when it captures stable structural invariants of the policy--outcome relationship. When policy interventions alter incentives, introduce thresholds, or change agent responses, transferred knowledge can become a source of misspecification rather than an advantage.

\section*{Code and Data Availability}

The experiments were generated using a Python notebook implementing the emissions-regulation simulator, the blank-slate piecewise learner, the restricted affine transfer learner, the sample-size experiment, and the online adaptation experiment. The simulation notebook, table-generation scripts, summary CSV files, and figure-generation code are provided as supplementary material. The notebook is self-contained and regenerates the reported CSV tables from the parameter settings reported in the paper.

\section*{Disclosure on the Use of Generative AI}

Large language model tools were used to assist with language editing, LaTeX formatting, figure-caption refinement, and code-debugging support. All scientific content, simulation design, numerical results, interpretation, and final claims were reviewed and validated by the author, who remains fully responsible for the manuscript.

\bibliographystyle{plainnat}

\end{document}